\documentclass[useAMS,usenatbib,epsfig, a4paper]{mn2e}

\usepackage{float}
\usepackage{epsfig}
\usepackage{times}
\usepackage{amssymb}
\usepackage{subfigure}
\usepackage{epstopdf}
\usepackage{verbatim}
\title[Large Scale Power from Peculiar Velocity Moments]
    {A Slight Excess of Large Scale Power from Moments of the Peculiar Velocity Field}
\author[E.~Macaulay et al.]
{E.~Macaulay$^1$\thanks{email: edward.macaulay@astro.ox.ac.uk}, H.~Feldman$^{2}$, P.~G.~Ferreira$^1$, M.~J.~Hudson$^{3,4,5}$, R.~Watkins$^{6}$\\ 
$^1$Astrophysics, University of Oxford, Denys Wilkinson Building, Keble 
Road, Oxford OX1 3RH, United Kingdom.\\
$^2$ Department of Physics and Astronomy, University of Kansas, Lawrence, KS66045, USA. \\
$^3$ Department of Physics and Astronomy, University of Waterloo, Waterloo, ONN2L3G1, Canada. \\
$^4$  Institut d'Astrophysique de Paris - UMR 7095, CNRS/Universit\'e Pierre et Marie Curie, 98bis boulevard Arago, 75014 Paris, France. \\
$^5$ Perimeter Institute for Theoretical Physics, 31 Caroline St. N., Waterloo, ON, N2L2Y5, Canada. \\
$^6$ Department of Physics, Willamette University, Salem, OR97301, USA. \\
}
\date{\today}

\pubyear{2010}

\begin{document}

\maketitle

\begin{abstract}

The peculiar motions of galaxies can be used to infer the distribution of matter in the Universe.  It has recently been shown that measurements of the peculiar velocity field indicates an anomalously high bulk flow of galaxies in our local volume.  In this paper we find the implications of the high bulk flow for the power spectrum of density fluctuations.  We find that analyzing only the dipole moment of the velocity field yields an average power spectrum amplitude which is indeed much higher than the $\Lambda$CDM value.  However, by also including shear and octupole moments of the velocity field, and marginalizing over possible values for the growth rate, an average power spectrum amplitude which is consistent with $\Lambda$CDM is recovered.   We attempt to infer the shape of the matter power spectrum from moments of the velocity field, and find a slight excess of power on scales $\sim1 h^{-1}$ Gpc.  \end{abstract}


\section{Introduction}

Peculiar velocities are useful cosmological probes. In principle, the peculiar velocity field is an unbiased tracer of the underlying matter distribution, and should be sensitive  to structure on scales larger than the nominal size of the survey.  The study of large scale flows has a long and hallowed tradition; throughout the 1980s and early 1990s, measurements of large scale flows were deployed to not only constrain the fractional energy density in matter,  $\Omega_m$ but also to identify possible sources of gravitational attraction which might lie outside current galaxy surveys (\cite{StraussWillick1995} ). 

Velocity catalogues are hard to construct. While current redshift surveys have of order $10^5$ objects, velocity catalogues are typically restricted to of order $10^3$ galaxies. Yet over the past three decades, enough measurements of peculiar velocities have been accrued to be able to construct a reasonably complete catalogue out to a maximum distance of about $100 h^{-1}$ Mpc. In a recent series of papers \citep{2009MNRAS.392..743W,2009arXiv0911.5516F}, the authors showed that a collection of peculiar velocity surveys could be combined to construct a reliable, well-behaved ``composite'' catalogue. This composite catalogue was then used to extract the simplest statistic: the bulk flow. The results were surprising: there is clear evidence for a large bulk flow, which is very unlikely within the current preferred model of structure formation, i.e. a flat, Friedman-Robertson-Walker universe with an appreciable cosmological constant permeated by Gaussian, scale-invariant perturbations (known as the $\Lambda$CDM model).
The findings of \cite{2009MNRAS.392..743W,2009arXiv0911.5516F} leads to the question: is there excess power on large scales? As yet we do not have a
direct measurement of the power spectrum of density fluctuations on these scales in the nearby Universe; the cosmic microwave background does probe a wide range of scales but at high redshift. Could the large bulk flow be a signature for large scale fluctuations at low redshift? It is this question we wish to address by using the measurement of the bulk flow and a few of the other lower moments to estimate the power spectrum of density fluctuations.  

\cite{1995ApJ...455...26J} inferred constraints on the matter power spectrum parameterized by $\sigma_8$ and shape parameter $\Gamma$ with data from \cite{lauer1994motion}, and seperately with data from \cite{1995ApJ...445L..91R}.  They found that including only the bulk flow of the Lauer \& Postman catalogue favoured a model with high large scale power, which became more reasonable when the shear of the velocity field was included, suggesting a very large scale excess of power.  However, the data from \cite{1995ApJ...445L..91R} did not suggest any large scale excess.

The  shear and dipole moments used by \cite{1995ApJ...455...26J}  are slightly different to the minimum variance shear and dipole moments we use here, because the \cite{1995ApJ...455...26J} method assumes that the non-modeled velocity is noise.  Thus the dipole moment from the \cite{1995ApJ...455...26J} method will depend on whether the shear is included, whereas the minimum variance formalism we employ here estimates each moment independently of the total number of moments analyzed.  Further, \cite{1995ApJ...455...26J} compare very different catalogues with different geometries, i.e., densities, distribution, etc.  The minimum variance method combines individual measurements to estimate the velocity moments for an idealized geometry, and the results are thus independent of the geometry of any particular survey.

\cite{1995astro.ph.12132K} inferred constraints on large scale structure from the POTENT velocity field.  Uncertainty in the growth of structure rate, $f$, was included by reporting the degenerate combination of $P(k) f^2$.  With current improved constraints on the growth rate, and more peculiar velocity data available, we attempt to constrain the unbiased matter power spectrum from moments of the peculiar velocity field.  

Recently \cite{2010JCAP...01..025S} applied a method to estimate bulk flows from two point correlations in galaxies distributions, and found no evidence for a large bulk flow in data from SDSS.  However, their analysis only probes scales up to 0.03 $h$Mpc$^{-1}$, whereas we see later that the anomalously high bulk flow is most sensitive to scales \textgreater 0.01 $h$Mpc$^{-1}$.

This paper is structured as follows.  In section \ref{combining_peculiar_velocities} we review the methodology to incorporate an individual peculiar velocity measurement into an estimate of our local peculiar velocity field.  We review the method of \cite{Kaiser_1988_bulk_flow} to weight peculiar velocity measurements to estimate the bulk flow.  We also review the method of \cite{2009MNRAS.392..743W} to use peculiar velocity measurements to estimate the bulk flow of our local volume, so that results from different surveys can be directly compared.  We  describe the peculiar velocity catalogues compiled by \cite{2009MNRAS.392..743W}. We then review the work of \cite{2009arXiv0911.5516F} to include higher moments of the velocity field.  In section \ref{velocity_to_matter} we describe how we relate these moments of the velocity field to constraints on large scale structure.  We present our results in section \ref{peculiar_velocity_results}.
\section{Combining Peculiar Velocity Measurements}
\label{combining_peculiar_velocities}

To estimate the peculiar velocity of galaxies at cosmological distances, we need both the measured redshift of the galaxy, and an independent distance measure, such as the luminosity distance.  From the distance measure, we can estimate what we would expect the redshift to be solely due to the Hubble flow.  We can then attribute the difference between this expected redshift and the measured redshift to the peculiar velocity of the galaxy (\cite{0691019339}).
There are two key difficulties with the method.  The first is that the uncertainty on the luminosity distance is typically rather large: $\sim10$ to $20\%$ for Tully-Fisher, Faber-Jackson, or Fundamental Plane measurements, and $\sim5\%$ for supernovae.  The second difficulty is more inherent: this method only provides the line of sight component of the peculiar velocity.  The importance of this effect can be illustrated if we consider a hypothetical bulk flow from an arbitrary north to the south.  A survey of galaxies to our north or south would be sensitive to this flow, whereas a survey to our east or west would not.  The approach we describe here is to combine individual peculiar velocity measurements to estimate the velocity field of our local volume by weighing each measurement according to how sensitive it is to each component of the underlying velocity field.
We can describe the velocity field ${\bf v}({\bf r})$, where ${\bf v}$ is the three components of the peculiar velocity field at position ${\bf r}$.  For each galaxy (labelled $n$), we measure the line of sight component of this field, $S$, i.e. $S_n({\bf r})={\bf v}({\bf r}_n) \cdot {\bf {\hat r}}_n$ where ${\bf {\hat r}}_n$ is the unit vector pointing in the ${\bf { r}}_n$ direction. We assume this measurement is drawn from a Gaussian distribution with variance $\sigma_n^2$.  An additional term, $\sigma_*^2 $, is included in the variance, to account for non linear flows.  The method is fairly insensitive to the particular choice of $\sigma_*^2$, as  the combined uncertainty of $\sigma_n^2 + \sigma_*^2 $ tends to be dominated by the measurement uncertainty, $\sigma_n^2 $.

The simplest result we can quote for a survey of peculiar velocities is the `bulk flow' vector, ${\bf u}$, the average velocity magnitude and direction of the galaxies in a survey.  We must be careful when combining peculiar velocity measurements to include the effect of only measuring the line of sight component.  This is achieved by multiplying each component of $S_n$ (in the coordinate system of the bulk flow) by a weight $w_a$, which essentially depends on the orthogonality of $S$ to the coordinate system of the bulk flow, so that $u_a=\sum_n w_{a,n} S_n$.  \cite{Kaiser_1988_bulk_flow} showed that the weights for the bulk flow are

\begin{equation}
w_{i,n}=A^{-1}_{ij}\sum_{m}{\frac{\hat{x}_j \cdot  \hat{r}_n}{\sigma_n^2+\sigma_*^2}}  
\label{bulk_flow_weights}
\end{equation}

where

\begin{equation}
A_{ij}=\sum_{m}{\frac{(\hat{x}_i \cdot  \hat{r}_m)(\hat{x}_j \cdot  \hat{r}_m)}{\sigma_m^2+\sigma_*^2}} 
 \label{A_ij}
\end{equation}

Because of sparse sampling, bulk flow results between different surveys are not necessarily immediately comparable.  \cite{2009MNRAS.392..743W} devised a method to weight the bulk flow of a particular survey to estimate the bulk flow of  a hypothetical spherically symmetric survey, where the density of galaxies $n$ falls of as $n(r)\propto exp(-r^2/2R^2_I)$, where $R_I^2$ is the characteristic depth of the survey.  This allows bulk flows from different surveys to be directly compared, and combined to give us a better estimate of the bulk flow of our local volume of space.  We start with the formalism from \cite{Kaiser_1988_bulk_flow}, and then introduce the method from \cite{2009MNRAS.392..743W}.

The idea is that the bulk flow we measure, $\bf u$, is essentially a convolution of the peculiar velocity field $\bf v$ with the window function of the survey $W$,
\begin{equation}
 u_i({\bf r_0})=\int d^3rW_{ij}({\bf r})v_j({\bf r+r_0})
  \label{u_window_function}
\end{equation}
This allows us to calculate the variance of $\bf u$ directly from the convolution theorem
\begin{equation}
 \left<u_{i} u_{j}\right>=\int d^3k P_v(k)W_{ij}^2(k)
   \label{u_variance}
\end{equation}
where the window function is given (in real space) by
\begin{equation}
W_{ij}(r)=A^{-1}_{ij}\sum_{n}\delta(r-r_n)\frac{\hat{r}_i\hat{r}_j}{\sigma_n^2+\sigma_*^2}   \label{tensor_window_function}
\end{equation}

It is extremely useful to split $\left<u_{i} u_{j}\right>$ into two terms: one term, $R_{ij}^{(v)}$, due to measurement variance, and a noise term, $R_{ij}^{(\epsilon)}$ due to the variance from non linear flows, so that $\left<u_{i} u_{j}\right>=R_{ij}^{(v)}+R_{ij}^{(\epsilon)}$.  We then have that 
\begin{equation}
R_{ij}^{(\epsilon)}=A_{ij}^{-1}
\label{error_matrix}
\end{equation}
and
\begin{equation}
R_{ij}^{(v)}=\int d^3k \mathcal{W}_{ij}^2(k) P_v(k)
   \label{velocity_matrix_geometry}
\end{equation}
where
\begin{equation}
\mathcal{W}_{ij}^2=W_{il}(k)W_{jm}^{*}(k)\hat{k}_l \hat{k}_m 
   \label{mathcal_W}
\end{equation}
We will return to equations \ref{error_matrix} and \ref{velocity_matrix_geometry} when we come to estimate large scale structure in section \ref{velocity_to_matter}.  

The approach of \cite{2009MNRAS.392..743W} is to find weights that minimize the average variance between a particular bulk flow vector $\bf u$ and the average bulk flow of our hypothetical survey, $\bf U$, $\left< (u_i-U_i)^2 \right>$.  If we assume the measurement error is uncorrelated with the bulk flow, we can expand $\left< (u_i-U_i)^2 \right>$ as
\begin{equation}
\left< (u_i-U_i)^2 \right> = \sum_{n,m}{w_{i,n}w_{i,m}  \left<S_nS_m \right>}+\left<U^2_i\right>  -2\sum_{n}w_{i,n}\left<U_iv_n \right> 
\label{expanded_variance}
\end{equation}
Before minimizing this expression with respect to $w_{i,n}$, we enforce the constraint that $\mathcal{W}^2_{ij}(k)\rightarrow 1/3$ as $k\rightarrow0$, so that the weighting is equal for each dimension of the peculiar velocity field.  We thus have to minimize
\begin{eqnarray}
& &\sum_{n,m}w_{i,n}w_{i,m}\left< S_n S_m \right> + \left< U_i^2 \right>\nonumber \\& & \ \ \ -2\sum_{n}w_{i,n}\left< u_i U_i \right>  
\lambda ( P_{nm}w_{i,n}w_{i,m}-\frac{1}{3}  )
\label{variance_to_minimize}
\end{eqnarray}
where $\lambda$ is a Lagrange multiplier and
\begin{equation}
P_{nm}=\int{\frac{d^2\hat{k}}{4\pi}\left(\hat{r}_n\cdot \hat{k} \hat{r}_m\cdot \hat{k} \right)}
 \label{P_nm}
\end{equation}

Differentiating equation \ref{variance_to_minimize}, and setting the result equal to zero gives
\begin{equation}
\sum_{m}(\left<S_nS_m \right>+\lambda P_{nm})w_{i,m}=\left<S_nU_i \right>  
\label{U_variance_non_matrix_form}
\end{equation}
  It is easier to solve for the weights, $w_{i,m}$, if we rewrite equation \ref{U_variance_non_matrix_form} in matrix form.  Substituting $\bf{G}$ for $\left<S_nS_m \right>$, $\bf{P}$ for $P_{nm}$, $\bf{w_i}$ for  $w_{i,m}$,  and $\bf{Q_i}$ for $\left<S_nU_i \right>$ gives
\begin{equation}
\bf(G+\lambda P){w}_{i}=Q_i
  \label{U_variance_matrix_form}
\end{equation}
which is now easy to solve for the weights $\bf{w}_{i}$. 
\begin{equation}
\bf{w}_{i}=(G+\lambda P)^{-1}Q_i
  \label{MV_weights}
\end{equation}

As well as the bulk flow of the survey, we can also consider higher moments of the peculiar velocity field if we consider the field as a Taylor expansion, as in
\begin{equation}
v_{i}({\bf r})=U_{i}+U_{ij}r_{j}+U_{ijk}r_{j}r_{k}+\dots
\label{velocity_field_taylor_expansion}
\end{equation}
$U_{i}$ is the bulk flow vector - also known as the `dipole moment'.  $U_{ij}$ is the `shear tensor' or the `quadrupole moment', and provides information about the distance at which the bulk flow attractor is located.  The `octupole tensor' $U_{ijk}$ provides information about the velocity field on scales smaller than the survey.  \cite{2009arXiv0911.5516F} have extended the work of \cite{2009MNRAS.392..743W} to include these higher order moments.
 If we assume that the peculiar velocity field is entirely due to gravitational infall, we expect the field to be curl-free, and consequently  $U_{ij}$ and  $U_{ijk}$ to be symmetric.  Thus the peculiar velocity field can be described to third order by the 19 independent velocity moments, $u_{i}$.

Before we proceed, we have to be careful with the definition of the octupole tensor, because some components of the tensor overlap with the dipole moment.  As such, we modify the expansion of the velocity field to be
\begin{equation}
v_{i}({\bf r})=U_{i}+U_{ij}r_{j}+U_{ijk}\left( r_{j}r_{k} - \Lambda_{jk} \right)+\dots
\label{velocity_field_taylor_expansion_minus_lambda}
\end{equation}
where $\Lambda_{jk}$ is given by
\begin{equation}
\Lambda_{jk}=\int_{V}r_jr_kd^3r
\label{lambda_for_velocity_field_taylor_expansion}
\end{equation}
in order to remove overlapping components.  The line of sight component is then
\begin{equation}
s({\bf r})=U_{i}\hat{r}_i+U_{ij}r\hat{r}_i\hat{r}_j+U_{ijk}\left(r^2\hat{r}_i\hat{r}_j\hat{r}_k - \Lambda_{jk}\hat{r}_i \right)+\dots
\label{line_of_sight_velocity_field_taylor_expansion}
\end{equation}
\cite{2009arXiv0911.5516F} have calculated the minimum variance weights for the 19 components of the third order expansion, over $k$ ranges from 0.002 to 0.196 $h$Mpc$^{-1}$.  
We can think of the resulting set of 19 components of the dipole, quadropole and octupole as a form of data compression containing the highest signal to noise information in a given peculiar velocity survey. Indeed, that will be our philosophy -
to use this form of the data to infer information about the underlying density field.

We study the `COMPOSITE' peculiar velocity catalogue compiled by  \cite{2009MNRAS.392..743W}, and also used in \cite{2009arXiv0911.5516F}. The catalogue consists of 4,536 peculiar velocity measurements, with a characteristic depth of 34 $h^{-1}$Mpc.  The characteristic depth is given by the average distance to each galaxy, weighted by the inverse square of the peculiar velocity uncertainty.

\section{Relating Velocity to Matter}
\label{velocity_to_matter}

We have presented a method to combine line of sight estimates of the peculiar velocity of individual galaxies into an estimate of the moments describing the velocity field of our local volume.  We now want to be able to compare this measurement to expectations from our cosmology.   The basic idea is that, in the linear regime, galaxies flow towards local over-densities of matter, so the velocity field 
${\bf v}({\bf r})$ is given by
\begin{equation}
 {\bf v}({\bf r})=\frac{f\,H_{0}}{4 \pi} \int d^3 {\bf r}' \delta({\bf r'})  \frac{({\bf r}'-{\bf r})}{\left| {\bf r}'-{\bf r} \right|^3 }
 \label{peculair_velocity_matter}
\end{equation}
$f$ is the perturbation growth rate, $\partial \ln \delta / \partial \ln a$, and $\delta$ is the matter density contrast.
The matter density contrast is modelled as a Gaussian random field with a power spectrum defined as $\langle |\tilde{\delta}(k)|^2\rangle=(2 \pi)^3 P(k)$, where $\tilde{\delta}$ is the Fourier transform of the density contrast in real space.

A useful way of parametrizing $P(k)$ is in terms of band powers:
\begin{eqnarray}
P(k)=\left\{\begin{array}{cl} P_\alpha &  k_\alpha < k < k_{\alpha+1} \\
0 & \mbox{otherwise} \end{array}\right.
\end{eqnarray}

To estimate the most likely matter power spectrum, based on the peculiar velocity data,  our approach is to construct, and then minimize, the likelihood function $\mathcal{L}$ as a function of $P_{\alpha}$:

\begin{equation}
-2\ln\mathcal{L}(P_{\alpha}) \propto \ln|\mathcal{C}|+{\bf u}^{T}\mathcal{C}^{-1}{\bf u}
\label{flow_log_likelihood}
\end{equation}
where ${\bf u}$ are the velocity moment components.  The covariance matrix, $\mathcal{C}$, is derived from equation \ref{u_variance}
\begin{equation}
\mathcal{C}_{pq}=R^{(v)}_{pq}+R^{(\epsilon)}_{pq}
\label{covariance_matrix}
\end{equation}
where the `error matrix' $R^{(\epsilon)}_{pq}$ is given by equation \ref{error_matrix}, and the `velocity matrix' $R^{(v)}_{pq}$ is given by equation \ref{velocity_matrix_geometry}.  We relate the velocity covariance matrix to the power spectrum by equation \ref{velocity_matrix_geometry}, so that
\begin{equation}
R^{(v)}_{pq}=\frac{f^2}{2 \pi^2}   \int dk P(k) \mathcal{W}^{2}_{pq}(k)
\label{velocity_matrix}
\end{equation}

We can choose the width of each band-power by integrating the window function (which is independent of $P_{\alpha}$) over the bin range $k_{\alpha}$ to $k_{\alpha+1}$.  We can then factor in the average power amplitude in this range, so that 
\begin{equation}
R^{(v)}_{pq}\simeq \frac{f^2}{2 \pi^2}  \sum_{\alpha}P_{\alpha}\mathcal{K}^{\alpha}_{pq}
\label{approx_velocity_matrix}
\end{equation}
where the kernel $\mathcal{K}$ is given by
\begin{equation}
\mathcal{K}^{\alpha}_{pq}=\int_{k_{\alpha}}^{k_{\alpha}+1} dk \mathcal{W}^{2}_{pq}(k)\
\label{kernel}
\end{equation}
We thus have a likelihood function for the velocity moments in terms of the growth rate and power spectrum band-powers.  

\begin{figure}
    \includegraphics[width=9cm]{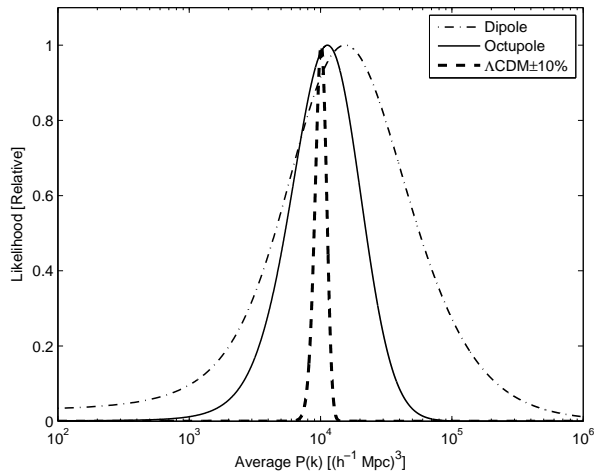}
\caption{Likelihood distributions for the average P(k) value in the $k$ range of our window function, inferred from a \emph{simulated} peculiar velocity catalogue.  Whether we include only the bulk flow of the velocity field (labelled `Dipole'), or also include the shear and octupole moments (labelled `Octupole'), we recover a value which is compatible with the fiducial $\Lambda$CDM value.  For clarity, the likelihood functions have been scaled arbitrarily to the maximum value of the dipole function.  The $\Lambda$CDM value has been illustrated here and in further such plots as a Gaussian distribution with an arbitrary standard deviation of $10\%$ of the fiducial value.}
    \label{simulated_average_pk_value}
\end{figure}

\begin{figure}
\centering
{
    \label{simulated_pk_shape}
    \includegraphics[width=9cm]{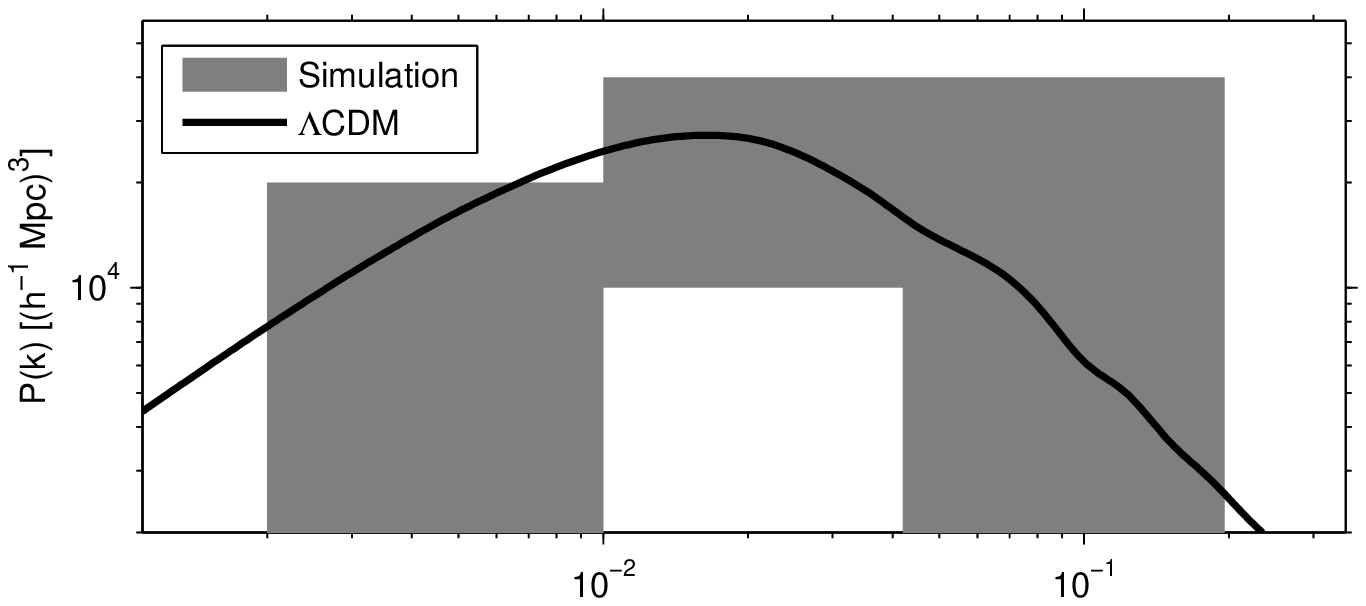}
}
\hspace{1cm}
{
    \label{window_function}
    \includegraphics[width=9cm]{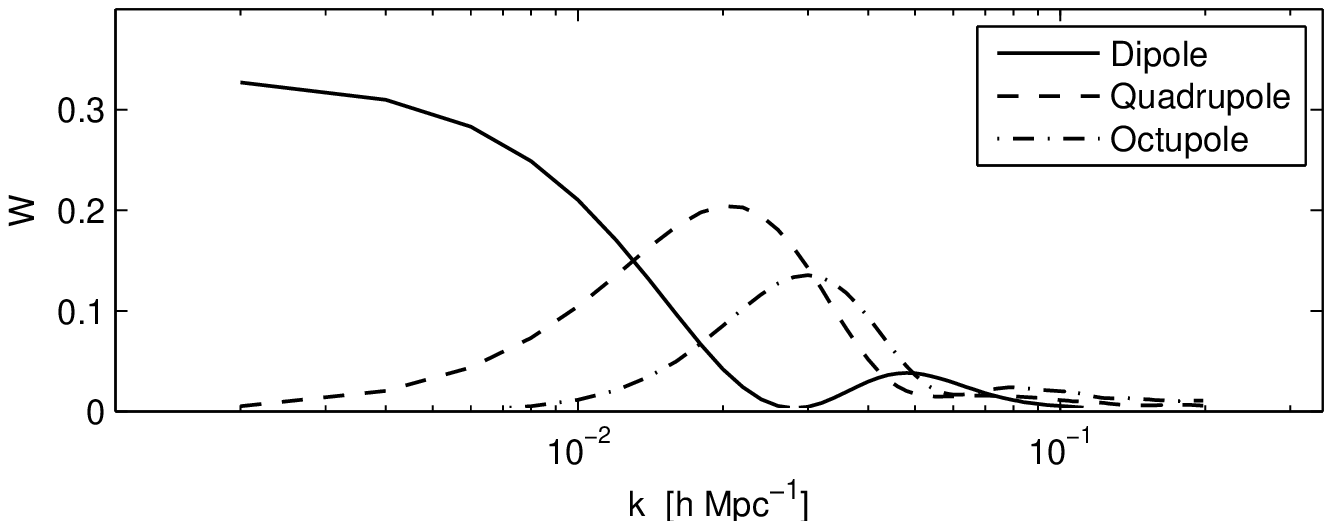}
}
\caption{In the upper figure is plotted the power spectrum shape, inferred only from peculiar velocity data generated from a simulated catalogue.  We have chosen three bands, evenly spaced in $\log k$.   The shaded region represents the marginalized $1\sigma$ uncertainty on the power in each band.  For comparison, the theoretical $\Lambda$CDM power spectrum has been overplotted.  The results shown here include the dipole, shear, and octupole moments of the simulated velocity field.  As discussed later, the uncertainties in each band have also been marginalized over the growth rate.  In the lower panel are plotted the average values of the dipole, shear, and octupole window functions.  The dipole is most sensitive on scales $k\lesssim$0.01 $h$Mpc$^{-1}$.}
\label{sim_pk_figure_with_win_function} 
\end{figure}

We have tested our methodology on simulated peculiar velocity catalogues, generated from a $\Lambda$CDM power spectrum.  We generate a series of mock catalogues from a set of N-body simulations, estimate the corresponding dipole, quadropoles and octopoles and then minimize the corresponding likelihood to recover the input power spectrum. We illustrate our results for one such realization in Figures \ref{simulated_average_pk_value} and \ref{sim_pk_figure_with_win_function}, and confirm that our method does not generate any spurious large scale power.

\section{Results and Discussion}
\label{peculiar_velocity_results}

\begin{figure}
\centering
\subfigure[ ] 
{
    \label{average_power_dipole}
    \includegraphics[width=9cm]{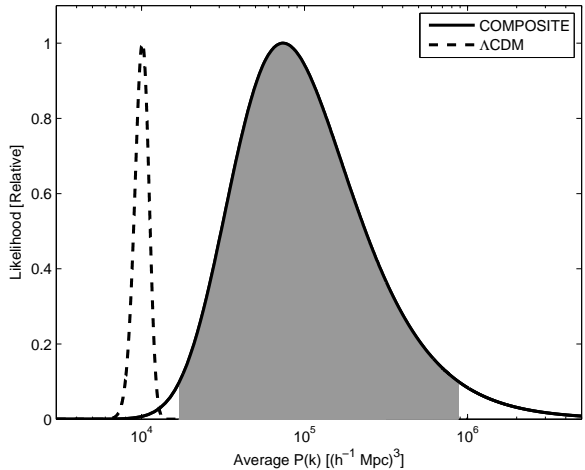}
}
\hspace{1cm}
\subfigure[ ] 
{
    \label{average_power_octupole}
    \includegraphics[width=9cm]{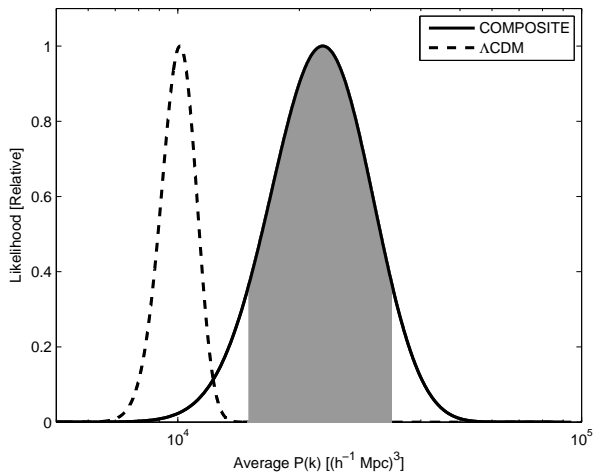}
}
\caption{In the upper panel is the likelihood distribution for the average amplitude of the power spectrum, inferred only from the dipole moment.    To account for the anomalously high bulk flow, we have to conclude a power spectrum amplitude which is incompatibly higher than the $\Lambda$CDM value.  In the lower panel, we show the effect of including the shear and octupole moments of the velocity field.    By now including the higher moments, the results are generally more compatible with the $\Lambda$CDM value.   Here, the growth rate, $f$, is kept fixed at the fiducial $\Lambda$CDM value.  In both figures, the shaded area is the 1$\sigma$ confidence region. The shaded area of the upper figure appears disproportionately wide because the large tail of the distribution is distorted by the logarithmic scale on the horizontal axis.}
\label{average_power} 
\end{figure}

We first consider the most likely total average power of our sample, over the entire $k$ range of our window function, $k=0.002$ to $0.196 h$ Mpc$^{-1}$.   We see in Figure \ref{average_power_dipole} that when we include only the dipole moment of the velocity field, the average power for each survey is much higher than the $\Lambda$CDM value.  This is entirely consistent with the results found in \cite{2009MNRAS.392..743W}.  When we also include the shear and octupole moments, we find much better agreement with the $\Lambda$CDM value, as we can see in Figure \ref{average_power_octupole}.  This is similar to the effect observed by \cite{1995ApJ...455...26J} when including the shear of the peculiar velocity field, and also noticed by \cite{2009arXiv0911.5516F}.

\begin{figure}
\centering
\includegraphics[width=9cm]{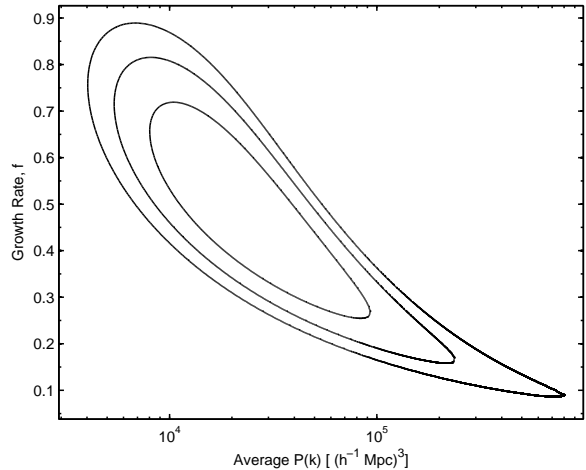}
\caption{Likelihood contours for the COMPOSITE survey, with dipole, shear, and octupole moments included, with contours spaced at 1, 2 and 3$\sigma$ confidence levels.  We are now considering the average power and the perturbation growth rate as free parameters.  A prior on the growth rate has been applied.  For comparison, the 1 dimensional likelihood distribution for the COMPOSITE survey in Figure \ref{average_power_octupole} is a slice through this likelihood contour at the fiducial $\Lambda$CDM value of the growth rate at $z=0$, i.e. $f=0.48$.  As can be seen here, including observationally constrained values of the growth rate allows for a wider range of allowed $P_{\alpha}$.
}

\label{composite_octupole_growth_power} 
\end{figure}

\begin{figure}
\centering
\includegraphics[width=9cm]{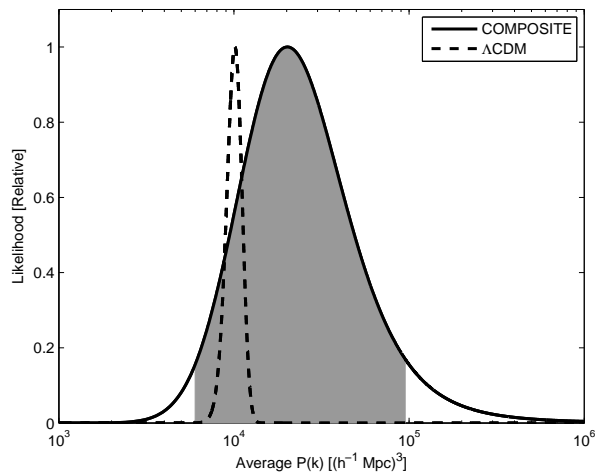}
\caption{Average power for the COMPOSITE survey, with dipole, shear, and octupole moments included, now marginalized over the growth rate.  The shaded region indicates the $1\sigma$ uncertainty boundaries on the likelihood function.  The expected $\Lambda$CDM value is now included in this range.}
\label{marg_average_power} 
\end{figure}

In Figure \ref{average_power}, the growth rate was fixed at the fiducial $\Lambda$CDM value, to illustrate the importance of including higher moments of the velocity field.  However, presently the best constraints on the growth rate at low redshift are from \cite{peacock2001measurement}'s measurement of the redshift space distortion compression parameter.  This constrains the growth rate to $f=0.49 \pm 0.14 $.  We now consider the effect of marginalizing over the growth rate, with this prior applied.  Likelihood contours for the growth rate and average power in the COMPOSITE survey are shown in Figure \ref{composite_octupole_growth_power}.  When we marginalize over the growth rate, we see in Figure \ref{marg_average_power} that the average power does not exclude the fiducial $\Lambda$CDM value. However in the context of $\Lambda$CDM models, the growth rate is not a free parameter, rather it is determined primarily by $\Omega_{m}$ (which also affects the shape of the power spectrum and the average power). In order to obtain consistency with the predicted average power, on requires $f \sim 0.7$, which is much larger than the fiducial $\Lambda$CDM value of 0.48.

In addition to the total average value, we can attempt to infer the shape of the power spectrum.  The window function was divided into three ranges, evenly spaced in $\log k$.  The likelihood function was then evaluated in terms of the power in each of the three bins, $P_{\alpha}$, and the growth rate.  As before, a prior on the growth rate was applied, and this parameter was marginalized.  When we only consider the bulk flow, we find that the likelihood is sufficiently broad and correlated that it is, in practice, impossible to pin down three independent estimates of $P_\alpha$. This is disappointing -- the measurement of the dipole alone does not allow us to identify if there is large scale power.

\begin{figure}
    \includegraphics[width=9cm]{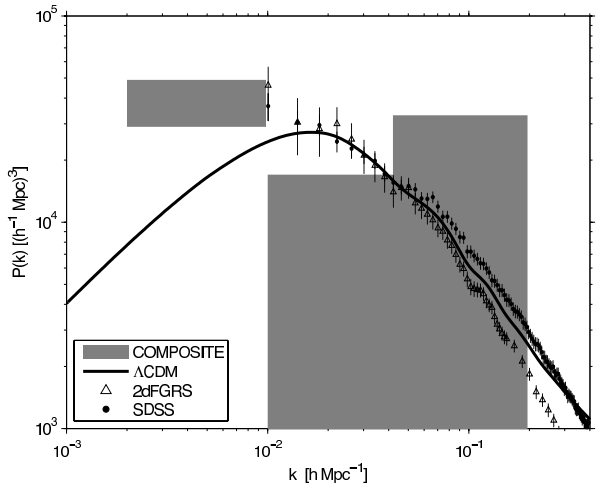}
\caption{Power spectrum shape inferred from dipole, shear, and octupole moments for the COMPOSITE catalogue.  The shaded regions are 1$\sigma$ uncertainties, marginalized over the growth rate and the other $P_{\alpha}$.   We find a slight excess of power on scales of  $\sim1h^{-1}$ Gpc.  These are the scales at which we are most sensitive.  This excess of power appears to agree with the largest scales in the 2dFGRS and SDSS surveys, which are plotted here as in \protect \cite{SDSS_Pk_shape} and have been deconvolved from their survey window functions.}
    \label{pk_shape_COMPOSITE}
\end{figure}

However, when we also include the shear and octupole moments of the velocity field, we find that the band powers  are virtually uncorrelated, and we can obtain three independent measurements.  In Figure \ref{pk_shape_COMPOSITE} we plot the marginalized 1$\sigma$ uncertainty for each $P_{\alpha}$ for the COMPOSITE catalogue.  There is a slight detection of excess power on the largest scales.  

We find that our estimate of the excess power seems to reflect the large scale
estimates of $P(k)$ found in the estimate from both the SDSS main galaxy power spectrum from \cite{SDSS_Pk_shape} and the 2dFGRS power spectrum from \cite{2005MNRAS.362..505C}. One should be careful about over-interpreting
these similarities, as both the SDSS and 2dFGRS estimates of power on large scales 
are, in principle, severely affected by edge effects. Indeed, the recent analysis
of the SDSS data in \cite{Sylos2009} should make one be wary about over interpreting estimates of structure on the largest scales of the survey. Yet the
fact that the same level of fluctuations is obtained from a peculiar velocity survey,
subjected to a completely different analysis may be an indication that redshift
survey estimates of structure on the largest scale must be taken seriously.  Furthermore, an excess of clustering on similar scales at higher redshift was recently found by \cite{2010arXiv1012.2272T} in the MegaZ DR7 photometric redshift survey.

What should we make of this excess of power, and how does it relate to our initial question: the anomalously high bulk flow?  To start with, we have demonstrated that it is incomplete to infer results from measurements of the bulk flow alone, as including higher moments of the velocity field yields significantly different results.  As noted in \cite{2009arXiv0911.5516F}, the low shear moments of the velocity field suggest that the sources responsible for the bulk flow are very far away.  In the formalism we have adopted here, only one of the bulk flow moments is anomalously high - the other bulk flow moments, and the shear and octupole moments, are consistent with $\Lambda$CDM.  From one perspective one could argue that the extra freedom from including 19 moments of the velocity field, and the growth rate, as opposed to just three moments of the dipole moment, provide a way to interpret the anomalously high bulk flow which is consistent with $\Lambda$CDM. 
On the other hand, all moments are not equivalent. The dipole moments are special in the sense that they probe the largest scales (and are also the moments that are most robustly measured).  Consequently the inferred shape of the power spectrum is different from $\Lambda$CDM.

\section{Acknowledgments}

We would like to thank Joe Zuntz and Andrew Jaffe for useful discussions.
PGF acknowledges support from STFC, BIPAC and the Oxford Martin School. HAF has been supported in part by an NSF grant AST-0807326 and by the University of Kansas General Research Fund.  MJH has been supported by NSERC and 
acknowledges the hospitality of the Institut d'Astrophysique 
de Paris, and the financial support of the IAP/UPMC visiting programme and the French ANR (OTARIE). 

\bibliographystyle{mn2e}
\bibliography{peculiar_velocity}

\end{document}